# The dawning of the *theory of equilibrium figures*: a brief historical account from the 17th through the 20th century

Giuseppe Iurato
*University of Palermo, IT*

**Abstract.** A brief but complete historical survey of the *theory of equilibrium figures* from its early origins, dating back to 17th-century, until the latest 20th-century developments, with a view towards its applications, is carried out.

**1. The early origins**

The *theory of equilibrium figures* of a self-gravitating rotating fluid mass investigates the existence and the stability of equilibrium solutions to the dynamical problem concerning the rotation of fluid systems. Sir A.S. Eddington once remarked[1] that one of most profound mysteries of the universe consists in the simple fact that everything rotates, by which one can easily understand why this theory is so important and widespread. In fact, it has played, and yet plays, a very fundamental role in many addresses of applied physics, ranging from relativistic astrophysics[2] to nuclear physics[3], although its theoretical framework has reached a satisfactory complete formal setting only in the late of 20th-century. In this paper, we shall only give a first brief historical survey of the main steps which have characterized the long route of this theory, in continuous evolution until up recent times. Following (Meinel et al. 2008, Preface), the theory of figures of equilibrium of rotating, self-gravitating fluids was developed in the context of questions concerning the shape of the Earth and celestial bodies. Many famous physicists and mathematicians, amongst whom are I. Newton, C. Maclaurin, C.G.J. Jacobi, J. Liouville, P.G.L. Dirichlet, R. Dedekind, B. Riemann, E.A. Roche, A. Lichtenstein, H.J. Poincaré, É. Cartan, and S. Chandrasekhar, made important contributions. Within Newton's theory of gravitation, the shape of the rotating fluid mass can be inferred from the requirement that the force arising from pressure, the gravitational force and the centrifugal force (in the corotating frame) be in equilibrium. According to (Roberts & Sousa Dias 1999, Section 1), in the *Principia*, Newton used his theory of gravitational attraction to show that an axisymmetric self-gravitating body of fluid that is rotating slowly about its axis of symmetry will be oblate, i.e., flattened at the poles. This result initiated a chain of mathematical discoveries spanning more than two centuries. Following (Lebovitz 1998, Section 1), it is possible to distinguish four main eras of the mathematical development of the theory of equilibrium figures, namely, an ancient history, including the discoveries made by Newton, Maclaurin and Jacobi; the dynamical equation period, due to Dirichlet, Dedekind and Riemann; the fission theory moment, including the work of Poincaré, Lyapunov and Cartan; and, finally, the recent development epoch, from the works of Chandrasekhar and his co-workers onwards. Herein, we shall consider only the first three main stages of this long and interesting historical pathway, since we are more interested in the early developments of the theory rather than in the modern and latest ones.

Following (Hildebrandt & Tromba 1996, Chapter 6), as has been said above, the theory of rotating bodies sprung out of celestial mechanics with first works by Newton, Maclaurin and A.C. Clairaut, though much before had already been made some observations which had led to the experimental evidence of the existence of rotating celestial bodies. Indeed, in 1611, the astronomer D. Fabricius, from his own astronomical data, reached the conclusion that the sunspots were integrant parts of the Sun itself and whose displacements were due to the Sun rotation, considering Sun have a spherical shape. In 1612, G. Galilei, T. Harriot and C. Scheiner, independently of each other, published the data of their own astronomical observations. In particular, Galilei agreed with Fabricius about the sunspots, whereas Scheiner considered these latter as a kind of little planets nearly orbiting around Sun. In his 1613 work *Istoria e Dimostrazioni Intorno alle Macchie Solari e Loro Accidenti*, Galilei refuted Scheiner arguments and, for the first time, officially supported the heliocentric theory of N. Copernicus. Later, Scheiner himself agreed with Galilei, and accomplished more precise and accurate measurements than those of Galilei, discovering that the Sun had a rotation period of about 27 days. Nevertheless, following (Appell 1932, Chapter I, Sections 2-3), (Lyttleton 1953, Chapter I), (Jardetzky 1958), (Chandrasekhar 1967), (Chandrasekhar 1969, Chapter 1) and (Lebovitz 1998), the study of the gravitational equilibrium of homogeneous uniformly rotating masses, formally began after having understood the importance of the gravitational law for the explanation of figures of celestial bodies, hence with the 1687 Newton's discussions of the figure of the Earth, in his celebrated *Philosophiæ Naturalis Principia Mathematica* (Book I, Section XIII and Book III, Sections XVIII-XX), where, once to be aware of the fact that the shape of the Earth would be spherical in absence of rotation, one would have wanted to predict the consequent departure from the spherical shape due to the centrifugal acceleration of the Earth's rotation. Thus, under the assumptions that the planet is a fluid mass of uniform density, the rotation is that of a rigid body, and the shape is that of an oblate spheroid

$$\frac{x^2 + y^2}{a^2} + \frac{z^2}{b^2}, \quad a > b$$

whose minor axis is the axis of rotation, Newton showed, anticipating many other hydrodynamical notions[4] and arguments which will be explicitated later, that the effects of a small rotation on the figure of the Earth, about an axis passing through the poles, must be in the direction of making it slightly oblate, the equilibrium of the body demanding a simple proportionality between the effect of rotation, as measured by the following parameter, called *ellipticity*,

$$\epsilon = (\text{equatorial radius} - \text{polar radius})/(\text{mean radius } R) \approx (a - b)/a,$$

and its so-called *cause*, as measured by the following parameter[5]

$$m = (\text{equatorial centrifugal acceleration})/(\text{surface mean gravitational acceleration}) = \Omega^2 R^3/GM,$$

where $G$ denotes the gravitational constant, and $M$ is the mass of the rotating body. Newton predicted $\epsilon = (5/4)m$, concluding that, if the Earth were homogeneous, then it should be an oblate spheroid with the following estimate for the ellipticity $\epsilon \cong 1/230$ (whilst today is known to be[6] $\sim 1/294$), having supposed, at that time, that $m = 1/290$. Nevertheless, Newton's predictions were quite contrary to the astronomical evidences of the time. At almost the same period[7], C. Huygens accomplished similar researches exposed in his famous 1690 *Discours de la cause de la pesanteur*, but pursuing another formal principle of the method in treating that. Following (Mach 1960, III.10.8), the mathematical hydrostatics began with these geodesic works of Newton, Huygens and Clairaut about the form of the Earth. In the meantime, research results due to G.D. Cassini and his school on the same question, were in contrast with the Newton's ones, because they provided an equatorial flattening rather than a polar one, but struggled in vain against the authority and reasoning of Newton. Thus, to try to clarify this controversy, P.L.M. de Maupertuis[8] and Clairaut himself organized a geodesic measurement expedition in 1738, which nevertheless confirmed Newton's prediction, with valuable credit to his new physics. Soon after[9], Clairaut wrote a little but important treatise, entitled *Théorie de la figure de la terre, tirée des principes de l'hydrostatique*, whose first edition appeared in Paris in 1743, a work that, according to Mach, inaugurated the rise of mathematical hydrostatics as a scientific discipline. In this work, Clairaut will give the first correct formulation of the Maclaurion's speculations. However, according to Lebovitz (1998), the above mentioned magnificent work of Newton was the first in a series of developments in the theory of Newtonian attraction that were to occupy some of the best mathematicians and physicists of Europe for many years to come.

---

[1] See (Jardetzky 1958, Foreword).
[2] See, for instance, (Tassoul 1978, 2000), (Cox 1980), (Shapiro & Teukolsky 1983), (Fridman & Polyachenko 1984), (Kippenhahn & Weigert 1990), (Padmanabhan 2000), and (Meinel et al. 2008).
[3] See, for instance, (Ring & Schuck 1980).

---

[4] Like that of fluid pressure.
[5] Which is a measure of the centrifugal acceleration in comparison to the mean gravity $G$.
[6] This discrepancy being likely due to the basic inhomogeneity of the Earth.
[7] See (Mach 1960, III.10.8).

**2. The contributions of C. Maclaurin and C.G.J. Jacobi**

The next valuable advance in the theory was owned to C. Maclaurin in 1742, who generalized Newton's outcomes to the case when the ellipticity, due to rotation, cannot be considered small. According to O. Struve (see Jardetzky 1958, Foreword)), in its early stages of the theory of rotating equilibrium figures, the main aim of the theoreticians was to trace the consequences of an ever increasing angular rate of rotation. Maclaurin gave a rigorous mathematical proof of the fact that an ellipsoid of rotation can be a figure of equilibrium of an isolated, rotating homogeneous fluid mass. To be precise, Maclaurin, after have solved an earlier problem concerning the attraction of an oblate spheroid at an internal point, deduced the following equation (*Maclaurin's formula*)

$$\frac{\Omega^2}{\pi G \rho} = \frac{(1-e^2)^{\frac{1}{2}}}{e^3} 2(3-2e^2) \sin^{-1} e - \frac{6}{e^2}(1-e^2)$$

being $\rho$ the density of the spheroid and $e = \sqrt{2\epsilon(1-\epsilon)}$ its eccentricity. With this formula, Maclaurin was the first to provide sufficient conditions for the existence of the hydrostatic equilibrium of a rapidly rotating fluid mass from which to determine the corresponding equilibrium surface. The discussion of this formula shows that there is a certain limit value of $\Omega$ and that, in general, at least three cases should be considered corresponding to the condition $h = \Omega^2/2\pi G\rho \lesseqgtr 0.2247...$, in the first two cases there being two possible *Maclaurin's ellipsoids* (or *spheroids*) corresponding to the same value of $\Omega$. If $h$ has exactly the above numerical value, or is larger, then only one out of such ellipsoids is possible. Finally, if $\Omega_0$ is the value of the angular velocity corresponding to the critical value $h \approx 0.2247$, then, for $\Omega > \Omega_0$, no ellipsoid of revolution may be a figure of equilibrium of a rotating fluid mass. Thereafter, starting from Maclaurin's formula, T. Simpson, in 1743, studied possible existence conditions for the Maclaurin's ellipsoids. He further realized that not ever a rapidly rotating mass will necessarily shape according to the figure of an oblate spheroid since, as $\Omega^2 \to 0$, we may have, by bifurcation (*d'après* Poincaré), two equilibrium solutions (the Maclaurin's spheroids), a first one which leads to a spheroid of small eccentricity (i.e., $e \to 0$), and a second one which leads to an highly flattened spheroid (i.e., $e \to 1$). From a historical viewpoint, it is generally believed that J.B. d'Alembert, in his celebrated and influential 1743 *Traité de dynamique*, was the first to explicitly notice this feature of Maclaurin's solution, even if Simpson gave before d'Alembert a simple table which distinctly implies this fact. Indeed Appell, in the third tome of his famous *Traité de mécanique rationelle*, states that this result will be correctly proved later by Laplace in 1776, following a remark due to d'Alembert.

---

[8] Who had made researches on this subject in 1732, in relation to planets.
[9] See (Kopal 1960, Chapter I).

For nearly a century after Maclaurin's discovery of two possible spheroids as equilibrium figures, it was believed that they represented the only admissible solution to the problem of the equilibrium of uniformly rotating homogeneous masses. This supposed generality of the Maclaurin's solution was never questioned, until up J.L. Lagrange, in his 1811 *Mécanique céleste*, considered formally the possibility of ellipsoids with unequal axes and satisfying the requirement of equilibrium, but concluding that at least two axes ought to be equal. Later, meanwhile C.G.J. Jacobi found, in 1834, a minor mistake in Lagrange reasoning, and reconsidering what had already been accomplished in previous potential theory works made by A.M. Legendre in the years[10] 1784-89, he pointed out that there may exist equilibrium figures which cannot be surmised from one may establish in the limit of spherical figures; for instance, ellipsoids with three unequal axes can very well be figures of equilibrium. The formal existence of these *Jacobi's ellipsoids* can be sketchily inferred by a simple extension of the Newton's original argument as follows. Indeed, at that time, it was known the components of the gravitational attraction, say $g_i$ $i = 1, 2, 3$, along the directions of the principal axes of an ellipsoid, say $a_i$ $i = 1, 2, 3$, can be expressed as $g_i = 2\pi G \rho A_i x_i$, where

$$A_i := a_1 a_2 a_3 \int_0^\infty \frac{du}{(a_i^2 + u)\Delta}$$

and $\Delta^2 := (a_1^2 + u)(a_2^2 + u)(a_3^2 + u)$. The formulas for the components of the attraction in the foregoing forms were first explicitly derived by C.F. Gauss in 1813 and by B.O. Rodrigues in 1815, independently of each other, in the context of potential theory, even if first forms of them may be also retraced in previous works of Legendre, Maclaurin himself as well as in a 1784 treatise of P.S. de Laplace, entitled *Théorie du mouvement et de la figure elliptique des planètes*. Remarkable contributions in this regard, were also provided by the work of J. Ivory in potential theory (see (Lebovitz 1998, Section 2)). Slightly modifying the original Newton's argument to the case of three axial ellipsoids, Jacobi stated, as sufficient conditions, the following relations

$$\frac{\Omega^2}{\pi G \rho} = 2 \frac{A_1 a_1^2 - A_2 a_2^2}{a_1^2 - a_2^2} = 2 a_1 a_2 a_3 \int_0^\infty \frac{u du}{(a_1^2 + u)(a_2^2 + u)\Delta}$$

and

$$a_1^2 a_2^2 \int_0^\infty \frac{u du}{(a_1^2 + u)(a_2^2 + u)\Delta} = a_3^2 \int_0^\infty \frac{du}{(a_3^2 + u)\Delta}$$

this last relation, for any assigned $a_1$ and $a_2$, allowing to get a solution for $a_3$ satisfying the following basic inequality

$$\frac{1}{a_3^2} > \frac{1}{a_1^2} + \frac{1}{a_2^2}$$

that, when $a_1 = a_2$, by means of the previous relations, determines a configuration common both to the spheroidal and the ellipsoidal sequences. A new limit value for the angular velocity, say $\Omega_1$, was determined for these Jacobi's ellipsoids; it is given by the condition $h \approx 0.1871$, so that when $h < 0.1871$, there is only one Jacobi's ellipsoid possible which corresponds to a given value of $\Omega$ and represents an equilibrium's figure. For the limit value $\Omega_1$ of the angular velocity, Jacobi's ellipsoid becomes an ellipsoid of revolution. Referring to this remarkable 1834 Jacobi's discovery, W. Thompson (Lord Kelvin) and P.G. Tait, in their 1867 *Treatise on Natural Philosophy*, refer that S.D. Poisson had already achieved similar results in the same period, even if, I. Todhunter, in his notable 1873 two-volume treatise *History of the Mathematical Theories of Attraction and Figure of the Earth from Newton to Laplace*, specifies that the Poisson's result was concerned with attraction of heterogeneous ellipsoids, whilst the Jacobi's one stated that an ellipsoid could be a possible form of relative equilibrium for a rotating fluid.

---

[10] Following (Appell 1926, Chapter I, Section 2), in these works Legendre, for the first time, used the powerful technique of harmonic function series expansions of the potential of the related dynamical system. The so-called *Legendre polynomials* (or *spherical harmonics*) will be introduced, for the first time, just in connection with these researches.

Nevertheless, in his short paper on this subject, Jacobi did not seriously examine the possible relationships between his ellipsoids with the Maclaurin's spheroids, this task having been pursued by C.O. Meyer in 1842, showing that the Jacobian sequence bifurcates[11] (d'après Poincaré) from the Maclaurin's sequence at the point where the eccentricity is $e = 0.81267$. This result can be easily obtained from the above relations by letting $a_1 = a_2$, so obtaining the following ones

$$\frac{\Omega^2}{\pi G \rho} = 2 a_1^2 a_3 \int_0^\infty \frac{u\,du}{(a_1^2 + u)^3 (a_3^2 + u)^{1/2}}$$

and

$$a_1^4 \int_0^\infty \frac{du}{(a_1^2 + u)^3 (a_3^2 + u)} = a_3^2 \int_0^\infty \frac{du}{(a_1^2 + u)(a_3^2 + u)^{3/2}}$$

where $\Omega^2/\pi G \rho$, on the left-hand side of the first relation of above, must now be identified with the one given by the above Maclaurin's formula. It can be show that both these last equations are simultaneously satisfied when $e = 0.81267$, where $\Omega^2/\pi G \rho \approx 0.37423$. Since it is known that the maximum value of $\Omega^2/\pi G \rho$, along the Maclaurin's sequence, is about $0.4493$, it follows that, for $\Omega^2/\pi G \rho < 0.37423$, there exist three possible equilibrium figures, to be precise, two Maclaurin's spheroids and one Jacobi's ellipsoid; for $0.37423 < \Omega^2/\pi G \rho < 0.4493$, only the Maclaurin's figures are possible; and, finally, for $\Omega^2/\pi G \rho > 0.4493$, no equilibrium figures are possible. This enumeration of the different possibilities is just due to Meyer. These results of Jacobi were better formalized by J. Liouville in 1834. Later, in 1846, Liouville restated Meyer's results using the angular momentum instead of the angular velocity, as the variable[12]. He further showed that while the angular momentum increases from zero to infinity along the Maclaurin's sequence, the Jacobi's figures are possible only for angular momenta exceeding a certain value, namely that at the point of bifurcation along the Maclaurin's sequence. But, the fact that no figures of equilibrium are possible for uniformly rotating bodies when the angular velocity exceeds a certain limit, raises the following issue: What happen when the angular velocity exceeds this limit? Whether some figures of equilibrium can exist or not if $\Omega$ surpasses the value $\Omega_0$, was an important problem first posed by P.L. Tchebychev in 1882 (see (Appell 1932, Chapter I, Section 2)). To be precise[13], in 1884 he told to A.M. Lyapunov about this his previous studies pursued upon the problem of the ring-shaped form of equilibrium of a rotating liquid mass whose particles are mutually attracted according to the

---

[11] Following (Michel 1975), this was the first example of a symmetry breaking phenomenon.
[12] Other contributions in this direction were also given by H. Smith in 1838 and G. Plana in 1853.
[13] See (Tchebychev 2008). For more information about *Tchebychev's problem*, see (Lyttleton 1953, Appendix) and references therein.

Newton's gravitation law. It is hard to know how far Tchebyshev advanced in this field, since he officially published nothing on this subject. Still, the very problem of the form of equilibrium of a rotating liquid mass, which he proposed to Lyapunov, was profoundly investigated by the latter who, together with A.A. Markov, was one of the Tchebyshev's most prominent pupils. Lyapunov was able to prove in 1884, making use of Legendre's methods, that there are no new figures of equilibrium in the neighbourhood of the limit ellipsoid $\Omega = \Omega_0$, but that there exist figures of equilibrium differing but little from some well-determined ellipsoids of Maclaurin and Jacobi type. Only one year later than Lyapunov, Poincaré too will find new kind of figures of equilibrium[14], similar to those found by Lyapunov, but with a different and independent method. In any case, from this point of the theory onwards, a prominent role gradually will be acquired by dynamical stability questions concerning the various possible figures of equilibrium, and, in this regard, Lyapunov's works will assume a central and pivotal role.

**3. The contributions of P.G. Dirichlet and G.B.F. Riemann**

In the last period of his life, namely in the winter of 1856-57, Dirichlet addressed himself to these questions, including the related topics in his lectures on partial differential equations in July 1857 but not publishing officially any detailed account of this his investigations during his life. The Dirichlet's paper, in which he found – albeit by means of an unfinished investigation – new equilibrium figures (*Dirichlet's ellipsoids*), was posthumously published by R. Dedekind in 1861. In this regard, Riemann wrote that Dirichlet opened up, in a most remarkable way, an entirely new avenue for investigations on the motion of a self-gravitating homogeneous ellipsoid. The precise problem that Dirichlet considered, consisted in investigating under what conditions one can have a configuration which, at every instant, has an ellipsoidal figure and in which the motion, in an inertial frame, is a linear function of the coordinates. Dirichlet formulated the general equations governing this problem in a Lagrangian setting, solving them in a detailed fashion for the case when the bounding surface is a spheroid of revolution, but he did not deeply investigate the figures of equilibrium admissible under the general circumstances of his formulation. Furthermore, the existence of ellipsoids of Maclaurin and Jacobi type was proved, by Dirichlet, as a special case of his theory. In this latter context, it was Dedekind, in an addendum to Dirichlet's paper, to explicitly prove a theorem that, nevertheless, Riemann said to be already implicitly present in the equations of Dirichlet's paper. Dirichlet's theorem concerns different possible motion's configurations which are related of each other by means of a suitable adjoint linear transformation of the reference frames. In particular, Dedekind considered the configurations which are congruent to the Jacobi ellipsoids and are their (functional) adjoints. These *Dedekind's ellipsoids*, while are congruent to the Jacobi ones, are stationary in an inertial frame and maintain their ellipsoidal figures by the internal motions which prevail[15].

The complete solution to the problem of the stationary figures admissible under Dirichlet's general assumptions was given by Riemann in 1861. He first shown that, under the restriction of motions which are linear in the coordinates, the most general type of figures of equilibrium compatible with an ellipsoidal one, consists of a superposition of a uniform rotation $\Omega$ with internal motions having a uniform vorticity $\zeta$ in the rotating frame. To be precise, Riemann showed that ellipsoidal figures of equilibrium (called *Riemann's ellipsoids*) are possible only under the following three circumstance, namely (*a*) the case of uniform rotation with no internal motions, (*b*) the case when the directions of $\Omega$ and $\zeta$ coincide with a principle axis of the ellipsoid, and (*c*) the case when the directions of $\Omega$ and $\zeta$ lie in a principal plane of the ellipsoid. The case (a) leads to the sequences of Maclaurin and

---

[14] Jardetzky (see (Jardetzky 1958, Chapter III, Section 3.1)) suggests more properly to speak of *Lyapunov-Poincaré figures of equilibrium*.

[15] Contributions to the stability questions raised by these Dirichlet's researches, were also provided by J. Hadamard in 1897.

Jacobi, the case (b) leads to the sequences of ellipsoids along which the ratio $f = \zeta/\Omega$ remains constant, the Jacobi's and Dedekind's sequences being special cases of these Riemann's sequences respectively for $f = 0$ and $\infty$, while the case (c), finally, leads to three other classes of ellipsoids. Riemann wrote down the equations governing the equilibrium of these ellipsoids and specified their domain of occupancy in the $(a_1, a_2, a_3)$-space. Riemann also sought to determine the possible stability conditions of these ellipsoids by an energy criterion[16], even if recently N. Lebovitz has found some minor mistakes in the related Riemann's argumentations. The problem of a varying homogeneous liquid ellipsoid was further investigated, amongst others, by F. Brioschi, R. Lipschitz, A.G. Greenhill, A.B. Basset, O. Tedone, E. Soler, P. Pizzetti, C. Mineo, A.E.H. Love, W. Steklov and R. Hargreaves. But, while Riemann's paper made an impressive starting urge towards the solution of Dirichlet's general problem, it left a large number of unanswered questions, amongst which, for instance, the one regarding the relation of the Riemann's ellipsoids with the Maclaurin's spheroids. Nevertheless, all these questions were to remain unanswered for more than a hundred years, mainly due to a spectacular discovery due to Poincaré which channelled all subsequent investigations along the directions which appeared to be rich of possibilities, above all on the astronomical side.

## 4. The contributions of H.J. Poincaré and A.M. Lyapunov

Thompson and Tait, in the 1883 third edition of their *Treatise on Natural Philosophy*, pointed out the necessity either to know what other possible figures of equilibrium there might exist between the Maclaurin's sequence and the Jacobi's one, and try to understand what happen beyond the latter sequence little by little rotation velocity increases. In this regard, the investigations relating to the equilibrium and the stability of the ellipsoidal figures of self-gravitating masses, for which Dirichlet and Riemann had laid such firm foundations, took an expected turn when Poincaré discovered[17], in 1885, that along the Jacobian sequence a point of bifurcation occurs similar to the one along the Maclaurin's sequence and that, even as the Jacobian sequence branches off from the Maclaurin's one, a new sequence of pear-shaped configurations (*Poincaré figures* or *pear-shaped figures*) branches off from the Jacobian sequence. Moreover, to warrant stability, Poincaré stated that the parameter $h$ couldn't exceed 1, a result that was later improved by U. Crudeli in 1910, who found an upper bound equal to $1/2$ (see (Appell 1932, Chapter I, Section 2)). In 1903, also on the basis of some remarks due to K. Schwarzschild in 1896, Poincaré[18] provided an estimate to the stability of a uniformly rotating steady-state configuration of the type $\Omega^2 \leq \pi G\rho$, where $\rho$ is the mean density. At almost the same time, G.H. Darwin provided as well other stability conditions for pear-shaped figures, in the years 1902-08. Besides what we have already said about Lyapunov's work in the previous sections, here we recall other his notable contributions to the subject, above in regard to stability problems (see (Appell 1932, Chapter I, Section 2)). From the 1880s onward[19], he addressed himself to the study and investigation of the theory of self-gravitating fluid systems. In 1884, he demonstrated, for the first time, that a sphere is the unique figure of equilibrium of an isolated fluid mass at rest; another proof of this fact, was later provided by T. Carleman in 1919. Lyapunov expressed his doubts about the possibility of proving the existence of new figures of equilibrium by using the method of successive approximations, so he developed another one consisting in finding a new figure of equilibrium which differs but little from an ellipsoid just comparing it, not to the given ellipsoid, but to a variable ellipsoid which is confocal to the former and passes through the point for which the value of the potential is considered. Lyapunov highlighted the various formal difficulties and inefficiencies of the previous methods used in dealing with the problem of finding equilibrium figures, so reworking out[20] a new and more correct power method essentially based on Legendre series expansion method of the potential[21]. On the other hand, the problem of figures of equilibrium becomes much more difficult if an inhomogeneous mass is considered, the first exact solutions in this direction having been provided just by Lyapunov in the early 1900s, even if some his results were only posthumously published. However, the very first attempts to lay and approach this hard problem were realized by Clairaut in his 1743 work mentioned above, so that one often speaks of the so-called *Clairaut's problem*. In this direction, outcomes to solve it, were pursued, amongst others[22], by H. Bruns, O. Callandreau, F.R. Helmert and M. Hamy between the 1870s and 1880s, as well as by P. Appell, P. Dive, H. Haalck, F. Hopfner, H. Jeffreys, W. Klussman, E. Soler, C. Mineo, P. Pizzetti, A. Véronnet, V. Volterra, E. Wiechert and R. Wavre, in the early 1900s.

---

[16] Almost contemporaneously, analogous works were attained by G.H. Bryan in 1889.

[17] On the basis of previous results achieved by P.L. Tchebychev (already quoted above), by Thompson and Tait in 1885 about a general astronomical issue, and by S. Kovalevsky in 1874 (but published only in 1885) when she was working, in Göttingen, on a problem concerning Saturn's rings from which arguments on the existence of a ring-shaped figure of equilibrium was possible to draw out. Similar researches concerning, above all, the shape and stability of the Saturn's rings and of some nebulae as well as the Sun's rotation, were accomplished by Laplace in the late 1700s, then by J.C. Maxwell and L. Matthiessen (in 1859), H. Faye (in 1880), N. Joukowski (in 1885), R. Emden (in 1907), B. Globa-Mikhailenko (in 1915), J.H. Jeans (in 1919), H. von Zeipel (in 1924), A.S. Eddington (in 1925), H.N. Russell and V. Bjerknes (in 1926), D. Riabouchinsky (in 1932), V.C.A. Ferraro (in 1937), C. Pekeris and Z. Kopal (in 1938), T.E. Sterne (in 1939), T.G. Cowling (in 1941), G. Randers (in 1942), J. Tuominen and W. Krogdahl (in 1944), and others (see (Jardetzky 1958), (Kopal 1960) and (Lebovitz 1965)). From all these works, the so-called *theory of zonal rotation* arose.

[18] See (Lebovitz 1967) and references therein.

[19] See (Jardetzky 1958, Chapter I).

**5. The contributions of L. Lichtenstein, G.H. Darwin, E.M. Roche and É. Cartan**

Following (Jardetzky 1958, Chapter V), the complexity and extreme length of Lyapunov's theory are due to the difficulties of the problem of equilibrium of a fluid mass and, of course, to desire not to leave out the proof of any point of the theory. A modification of the method of Lyapunov was given by L. Lichtenstein in a series of papers and works which started from 1918 until up the basic 1933 treatise entitled *Gleichgewichtsfiguren rotierender Flüssigkeiten*, a complete synthesis of the previous Lichtenstein's works on this subject, in which many results achieved by Poincaré and Lyapunov could be obtained above all through the resolution of suitable integral equations. Lichtenstein, moreover, provided many other new results concerning the theory of equilibrium figures as, for instance, those regarding symmetry questions. Amongst the latter, we mention a notable condition that a rotating mass must satisfy and that it is expressed by an important theorem, for the first time given by Lichtenstein in 1918, according to which, for a rigid-body rotation and uniform density, the related uniformly rotating steady-state configuration must have a plane of symmetry perpendicular to the axis of rotation; this result was then extended to angular velocities independent of $z$, as well as to a wide class of density stratification. Lichtenstein also supplied a method for treating inhomogeneous problems, extending parts of the Lyapunov's theory, even to be able to provide new equilibrium figures for a class of heterogeneous masses. On the other hand[23], the Lyapunov's method is still able to treat the small changes in the figure of a rotating fluid mass due to the existence of some other bodies. In the problem of the figure of a liquid planet which is affected by the attraction of other members of a dynamical system of solar system like, the most precise method was provided by Lichtenstein. In this regard, ever since 1848 É.A. Roche dealt with the problem (said to be the *classical Roche problem*) concerning the figure of a homogeneous liquid mass subjected to the attraction of a far-removed mass centre, this problem having been treated formally by P. Appell in 1920s and later quite generalized by C. Agostinelli in 1940s. Following (Lebovitz 1965), roughly a *Roche model* consists of a core containing all the mass $M$, producing the gravitational potential $GM/r$ into an overlying atmosphere of negligible mass, the level surfaces being those for which $U$ is constant. The main dynamical problem thus consists in studying the dynamical evolution of a sequence of such Roche models (*Roche's sequence*), labelled by the angular momentum.

---

[20] These Lyapunov's efforts were pursued since the late 1800s, but notwithstanding that they were published, in a final form, only posthumously in 1925, after more than 30 years of investigations in this field. Other improvements of the Lyapunov's work were attained, amongst others, by U. Crudeli, P. Humbert and M. Orlov in the early 1900s.
[21] This method of series expansion of the potential will be then improved and enlarged by R. Wavre (with his *uniform method*), as well as by C. Mineo, U. Crudeli and P. Dive, in the early 1900s.
[22] See (Berzolari 1975, Section L.I.1), (Appell 1932, Chapter I, Section 3) and (Kopal 1960).
[23] See (Jardetzky 1958, Chapter X).

The 1885 result of Poincaré is equivalent, in modern terminology[24], to the statement that, along the Jacobian sequence, there is a point where the ellipsoid allows a neutral modes of oscillation belonging to the third zonal harmonic, while a corollary, also enunciated by Poincaré, states that, along the Jacobian sequence, there must be further points of bifurcations where the Jacobian ellipsoids allow neutral modes of stellar oscillation belonging to the fourth, fifth and higher zonal harmonics. Poincaré stated as well other notable conjectures related to astronomical questions, his legacy having been profitably reconsidered and deepened by many other scholars who spent years of efforts toward the substantiation of these conjectures. Nevertheless, no serious further attempt was pursued to investigate the stability of the Maclaurin's spheroids and the Jacobi's ellipsoids from a more proper direct analysis of normal modes. Finally, É. Cartan[25], in 1924, established that Jacobi's ellipsoids become instable at its first point of bifurcation and behaves, in this respect, differently from the Maclaurin's spheroids which, in the absence of any dissipative mechanism, is stable on either side of the point of bifurcation where the Jacobian sequence branches off. But an another important problem of related interest in the theory of ellipsoidal figures of equilibrium was formulated by Roche in the years 1847-50: i.e., he considered the equilibrium of an infinitesimal satellite (of density $\rho$) rotating about a rigid spherical planet (of mass $M'$) in a circular Keplerian orbit (of radius $R$), and showed that no equilibrium figures are possible if the angular velocity ($\Omega$) of orbital rotation exceeds the limit $\Omega^2/\pi G\rho = M'/\pi\rho R^3 \leq 0.090093$. The lower limit to $R$ set by the foregoing inequality, is called the *Roche limit*. Roche also considered the case when the mass of the satellite is finite, and showed that inequalities analogous to the last one, exist. But in all of Roche's considerations, the assumption of a rigid spherical body for the distorting mass was retained. In 1906, Darwin, with a view toward application to double stars, attempted to allow for the mutual distortion of the components by an approximate procedure, but his efforts were only partially successful. An interesting case, when in Roche's problem only the tidal forces are taken into account, was considered by Jeans in 1906. The equilibrium and stability of the ellipsoids of Jeans, Roche and Darwin, form a separate consistent chapter in the theory of ellipsoidal figures of homogeneous masses.

At this point, we are achieved a position facing the modern mid-1900s developments of the theory, mainly due to Chandrasekhar and his school, as well as by other scholars such as R.A. Lyttleton, H.P. Greenspan, Z. Kopal, W.S. Jardetzky, and many others, mainly working on astronomy and celestial mechanics. The current developments of the theory will make use of modern and sophisticated mathematical and mathematical-physics tools and techniques of various nature, like those of the theory of stellar pulsation[26], methods and techniques of perturbation theory, group theory[27], variational calculus[28], dynamical system theory[29], and so forth. But, to give an although brief account of the latest developments, it would need another paper as the present one.

**Acknowledgements.** It has been thanks to the kind courtesy of Professor James Montaldi that I have become aware of the fact that also dynamical system theory has interesting and fruitful applications to the theory of equilibrium figures. He also has gently provided me some bibliographical hints in regard to the these latter applications of dynamical system theory whose initial motivation was to use methods from Geometric Mechanics, i.e., using symmetry and conservation laws as the organizing and methodological principle on the wake of what Riemann originally did.

---

[24] Referring to the harmonic series expansion of the potential of the gravitating fluid mass.

[25] Often, instead to correctly quote Élie Cartan, strangely enough many textbooks on the subject-matter wrongly refer to his son Henry Cartan.

[26] See, for instance, (Cox 1980), (Fridman & Polyachenko 1984), (Tassoul 1978; 2000) and (Meinel et al. 2008).

[27] See, for instance, (Constantinescu et al. 1979) and references therein.

[28] See, for instance, (Hildebrandt & Tromba 1996).

[29] See, for instance, (Roberts & Sousa Dias 1999), (Fasso & Lewis 2001), (Rodriguez-Olmos & Sousa Dias 2009) and references therein.